**Maximizing absorption in photon trapping ultra-fast silicon photodetectors**

*Cesar Bartolo-Perez, Wayesh Qarony, Soroush Ghandiparsi, Ahmed S. Mayet, Ahasan Ahamed, Hilal Cansizoglu, Yang Gao, Ekaterina Ponizovskaya Devine, Toshishige Yamada, Aly F Elrefaie, Shih-Yuan Wang, M. Saif Islam\**

C.Bartolo-Perez, Dr. W.Qarony, Dr. S.Ghandiparsi, A.S.Mayet, A.Ahamed, Dr. H.Cansizoglu, Dr. Y.Gao, Prof. M.S.Islam
Electrical and Computer Engineering, University of California—Davis, Davis, California, 95618, USA
E-mail: sislam@ucdavis.edu

Prof. T. Yamada
Electrical Engineering, Baskin School of Engineering, University of California, Santa Cruz, California, 95064, USA

Dr. E. P.Devine, Dr. A.F.Elrefaie, Dr. S.Y.Wang
W&WSens Devices, Inc., 4546 El Camino, Suite 215, Los Altos, California, 94022, USA

Silicon photodetectors operating at near-infrared wavelengths with high-speed and high sensitivity are becoming critical for emerging applications, such as Light Detection and Ranging Systems (LIDAR), quantum communications, and medical imaging. However, such photodetectors present a bandwidth-absorption trade-off at those wavelengths that have limited their implementation. Photon trapping structures address this trade-off by enhancing the light-matter interactions, but maximizing their performance remains a challenge due to a multitude of factors influencing their design and fabrication. In this paper, strategies to improve the photon trapping effect while enhancing the speed of operation are investigated. By optimizing the design of photon trapping structures and experimentally integrated them in high-speed photodetectors, a simultaneous broadband absorption efficiency enhancement up to 1000% and a capacitance reduction of more than 50% has been achieved. Such work also allows to present empirical equations to correlate the quantum efficiency of photodetectors with the physical properties of the photon-trapping structures, material characteristics, and limitations of the fabrication technologies. The results obtained, open routes towards designing cost-effective CMOS integrated.

## 1. Introduction

Conventional silicon photodetectors (PDs) have weak absorption capabilities at near-infrared (NIR) wavelengths, forcing them to be designed with thick absorbing layers to obtain high efficiency at the expense of limited bandwidth of operation. Hence, the trade-off between bandwidth and efficiency limits the use of Si, while other alternatives like gallium arsenide (GaAs) are expensive and incompatible with CMOS technologies. Considerable efforts have been devoted to enhancing the efficiency and bandwidth simultaneously in Si-based photodetector to utilize the cost-effective and matured CMOS fabrication capabilities.[1-4] Detecting low levels of light at NIR at high bandwidth is critical in emerging applications like LIDAR, where a 3D image of the environment is required for autonomous vehicles, virtual/augmented reality systems, and robotics.[5-8] PDs with high signal-to-noise ratio also allow building efficient optical communication systems,[9-11] while highly sensitive and fast PDs facilitate high-resolution biosensors, spectroscopy,[12] and medical imaging technologies.[13-15] In this effort, several photon-trapping (PT) structures,[4, 16-23] which have been successfully implemented in solar cells, are now being implemented in PDs.[24-27] By introducing such PT structures on the surface, the reflection of light is reduced and the optical path length is increased, enhancing the photon absorption and allowing the use of thinner semiconductor layers for faster carrier collection.[1, 28] Using this approach, ultrafast and highly sensitive PDs have been demonstrated by several groups both for short [24, 25, 29-31] and longer wavelengths.[26, 32-35] However, due to the different degrees of freedom for design and fabrication, the implementation of PT structures in PDs remains a challenge. Extensive design variations can be analyzed by numerical methods to optimize the performance of the device.

This extensive exercise along with the uncertainties in the fabrication processes contribute to significant challenges in optimizing the device performance.

We conducted extensive simulation and designed vertical pin photodetectors with more than 150 unique integrated PT structures by varying size, shape, period, and orientations, and established a crucial correlation between these parameters to enhance the device performances. Our rigorous simulations and extensive experimental investigations enabled a combination of optimum parameters to help to overcome the trade-off between bandwidth and efficiency in the PDs. Besides, it allows high sensitivity for low levels of photon detection with 500% higher external quantum efficiency (EQE) as compared with the conventional PDs at 850 nm wavelength, and up to 1000% enhancement at other NIR wavelengths around 1000 nm. Additionally, our fabricated devices exhibit more than 50% reduction in junction capacitance due to the introduction of PT structures, and this, in turn, improves the device bandwidth. The extensive design variations make it the most comprehensive study aimed at understanding the PT phenomenon in high-performance PDs. To enable performance projections, it is of interest to develop simple, closed-form expressions for the EQE of a high-speed PT photodetectors that intuitively connect the physical parameters of the PT structures, material characteristics, and quality of fabrication. This work elucidates such crucial expressions to enable the implementation of the PT structures for absorption efficiency enhancement, capacitance reduction, and faster time response.

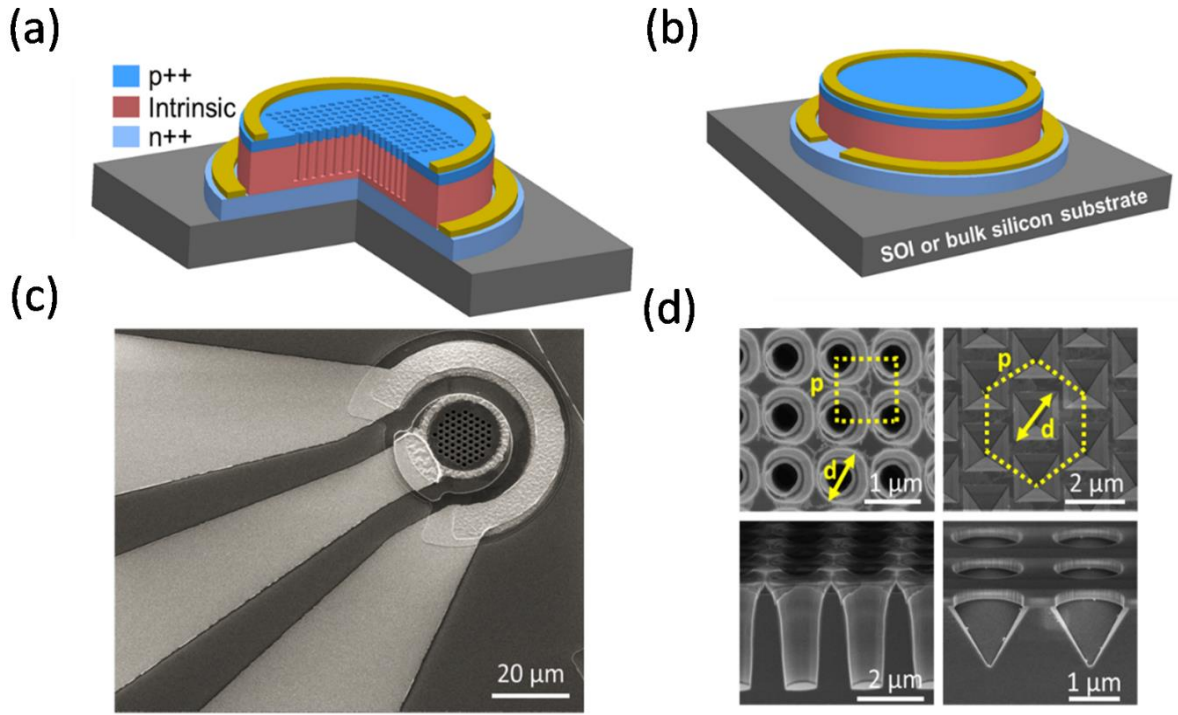

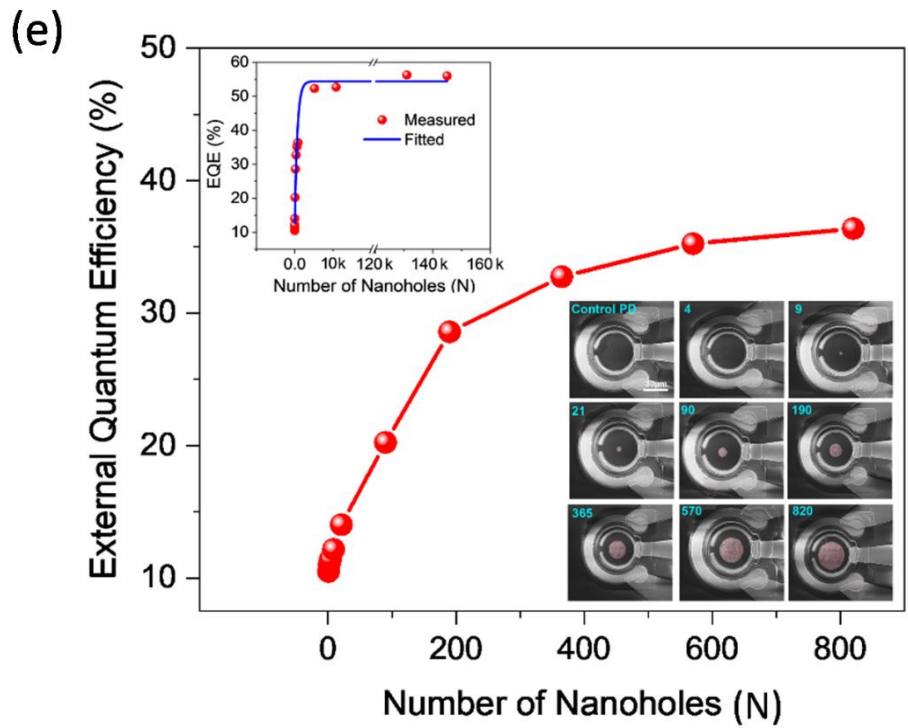

**Figure 1**. a) Schematic of Si pin photodetector (PD) with integrated PT structures for an i-layer of 2 μm thickness. b) A control PD without PT structures. Both devices are fabricated on SOI and bulk Si substrate. c) SEM image of PT PD with high-speed coplanar waveguide. d) Top and cross-sectional views of PT structures with funnel and inverted pyramid shapes for both square and hexagonal lattices. Over 40 different PT design variations in diameter, period, and shape are investigated. e) EQE versus number of nanoholes (N) in PDs under 850 nm laser illumination. Insets: (top) fitted curve shows that EQE starts with ~12% efficiency for control PD and saturates at ~56% for PT PD with ~5000 nanoholes. (bottom) SEM images of PDs with an increasing N in devices with 50 μm diameter.

## 2. Results and Discussion

### 2.1. Device design and Fabrication

The cross-sectional designs of mesa type Si pin PDs with and without integrated PT structures are schematically shown in **Figure 1**a-b, respectively. The PDs consist of heavily doped p (p++) and n (n++) type Si layer and a 2 μm of thin intrinsic layer, epitaxially grown on the top of Silicon on Insulator (SOI) or bulk Si substrates. The PDs are fabricated with a different diameter (D) ranging from 30 to 500 μm. An SEM image of fabricated PD is depicted in Figure 1c. An array of micro/nanoholes is patterned with a funnel shape or inverted pyramid etching profiles on the surface of the PDs, which serves as potential PT structures. These nanoholes are distributed in hexagonal or square lattices, designed with different diameters (d) and periods (p) ranging from 630 to 1500 nm and 900 to 3000 nm, respectively. Figure 1d represents the top and cross-sectional SEM images of PT holes with a tapered funnel and inverted pyramid shapes. The dimensions of the PT structures are selected to be close to the wavelengths of interest. The periodic distance between structures is reduced in each set of devices by keeping the d fixed, allowing to increase the number of nanoholes that can be accommodated on the surface of PD. The depth of the nanoholes was etched to be around 2 μm for funnel shape, while it varies between 450 to 1000 nm for inverted pyramids. Also, an unpatterned device is fabricated as a reference which we call a control device to compare with the PT PDs (See Figure S1 and Table S1 in Supporting Information for more details on the investigated devices with the variation of different parameters, such as substrate, the diameter of PT holes, and periods).

## 2.1. External Quantum Efficiency

EQE is one of the key parameters that describes the sensitivity of PDs. Herein, several PT PDs with integrated nanoholes are designed, optically simulated, and fabricated, while the period, diameter, and number of nanoholes ($N$) are systematically varied. Employing the following design guidelines, one can easily fabricate an optimized device by performing simulations beforehand. Firstly, the influence of the $N$ on the measured EQE is studied, where a set of devices with a constant device $D$ of 50 µm is characterized. SEM images of such devices with different N are illustrated in the inset of Figure 1e and Figure S2. The measured EQE is presented in Figure 1e for PT PDs with a fixed period and diameter of 1000 and 700 nm, respectively for hexagonal lattice and inverted pyramid profile on SOI, where the N is varied from 0 (control) to 820. Compared to the control device with an EQE of ~12%, the EQE of the PT devices gradually increases with increasing $N$, exhibiting a maximum of >38% for an N value exceeding 820. It is important to note that the test devices above were not among the designs with optimum parameters. This experiment mainly demonstrates a correlation between the EQE of a photodetector and N. Other periods and diameters that were optimized contributed to considerably higher peak efficiencies. The EQE enhancement observed in the device is due to the improved coupling of vertically incident light into laterally propagating modes with increasing $N$ within the same area of the devices. Besides, a reduction of planar area in a device leads to decreased surface reflection and improved transmission of the incident light, resulting in relatively higher absorption in the photoactive layer. Consequently, the overall EQE of the PT devices is distinctly increased in comparison to the control device.

Next, EQEs of 500 μm devices with higher $N$ values and maximum up to 145000 nanoholes with the same design as 50 μm diameter devices are added to establish a relationship as shown in Figure 1e (top inset). It shows that, for this design with $d/p$=700/1000, the EQE value can saturate at ~56% for approx. 5000 nanoholes. The maximum $N$ presented in Figure 1e is 820 with a filling fraction (Area$_{holes}$ / Area$_{device}$) of only 16% for the 50 μm device (see Supporting Information Figure S3 for other filling fraction values), while a maximum of about 5000 nanoholes can be accommodated in the same device contributing to a very high filling fraction. Advanced foundry processes can accommodate almost 100% filling fraction by reducing the size of the features (such as contact electrodes, the separation between the region covered by the holes, and interconnect) using tighter fabrication tolerances.

Based on the fitting curve, an empirical equation can correlate the EQE of the devices with the photon-trapping structural and device parameters,

$$\eta_{PT} = \eta_{max} \times \Delta - a\Delta \times \left(\frac{p}{d}\right)[exp\,(-bN)] \quad (1)$$

where $\eta_{PT}$ is the EQE of a PT PD for a specific $N$, $\eta_{max}$ is the maximum possible EQE (simulated value) for the device, a and b are design constants which were calculated to be 41 and 0.00147, respectively, for this design, and $\Delta$ is the ideality factor of the device. $\Delta$ represents the degrees of perfection in the fabrication process and material quality. When the value of $\Delta$ is 1, $\eta_{PT}$ value gets closer to $\eta_{max}$. Imperfection in device fabrication and the impurity of materials could lead to $\Delta$ smaller than unity, whereas in our devices, it varied between 0.75 to 0.81. Several of the $\eta_{PT}$ values of fabricated devices were calculated based on Equation 1 with varying $N$, where a good agreement between the calculated values and the performances of the fabricated devices was observed (see Table S3, Supporting Information). Equation 1 is valid for any device diameter and can accommodate any number of nanoholes, as shown in the inset 1e.

Next, the device is further optimized by varying etching profile, d, p, and substrates (SOI and Si bulk substrate) to maximize the efficiency. The investigated PDs with 500 μm diameter have a filling fraction as high as ~45%. The EQE of such PDs measured as a function of *d/p* is depicted in **Figure 2**a, where the EQE values increase almost in a linear fashion as *d/p* changes from 0.4 to 0.8 due to the different 2D hole crystalline symmetries. The measured EQE reaches its maxima at *d/p* = ~0.8, which is due to the maximum influence of photon interaction with the 2D hole array, and equivalently the slowest photon velocity or the maximum photon trapping, as illustrated in Figure S5 of Supporting Information. The maximum EQE exhibited by the control device is about 15%, which is in good agreement with the FDTD simulated control device. The PT PDs fabricated on a bulk Si wafer exhibit EQEs between 15% and 25%. However, compared to the control and PT devices fabricated on bulk Si, the EQEs of PT PDs on SOI substrate distinctly increased, resulting in a very high EQE ranging between 30% and 56%. Subsequently, the influences of the etching profile of PT structures on the EQE are studied. The PT PDs fabricated on SOI substrate are arranged in hexagonal or square lattices with an inverted pyramid or funnel shape etching profiles. PDs with an inverted pyramid exhibit relatively higher EQE than the devices with a funnel shape. For instance, the inverted pyramid PDs with d/p of 0.81 exhibit an EQE over 60%, whereas the EQEs exhibited by funnel shape are lower than 50% (see Figure S4, Supporting Information). This discernible enhancement obtained in the inverted pyramid can be attributed to the effective refractive index gradient in the interface of air and Si, resulting in a superior antireflection effect and efficient coupling of light over a wide wavelength and angular ranges.[36] This is an added advantage of PT structures over traditional quarter-wavelength thin-film antireflection coatings.[37]

The experimental results can lead to an empirical equation to capture the correlation between $\eta_{PT}$ and the d/p in a very generalized fashion,

$$\eta_{PT} = \eta_{flat} + \left(\frac{d}{p}\right) \Delta \times \beta \quad (2)$$

where $\eta_{flat}$ is the EQE of the control devices and $\beta$ is the PT factor. $\beta$ can be determined from the slope of a linear curve connecting multiple EQEs as a function of $d/p$. Equation 2 is valid for PD with both hexagonal and square lattices with varying $\beta$ values, provided that a sufficient number of PT nanoholes are integrated on the surface of the PDs to reach saturation level in photon absorption (see Table S2, Supporting Information). Prior to fabrication, one can determine $\beta$ by simulating a set of PDs as a function of $d/p$. The linear fitting curve drew using Equation 2 for the simulated funnel shape PDs is shown in Figure 2a. Several $\beta$ values and $\eta_{PT}$ for the simulated and fabricated devices are provided in Table S4 and S5 and are found to be in good agreement with our observations. With a set of ideal simulations, one can estimate the efficiency of the fabricated devices as a function of $d/p$ by following this design guideline.

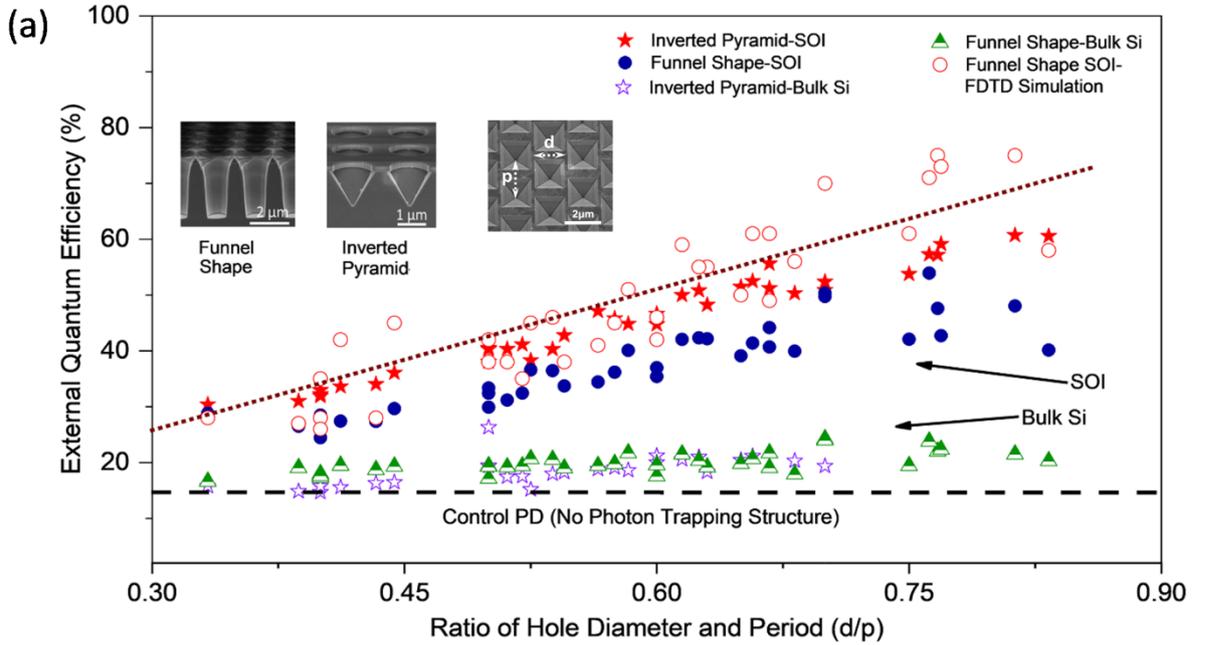

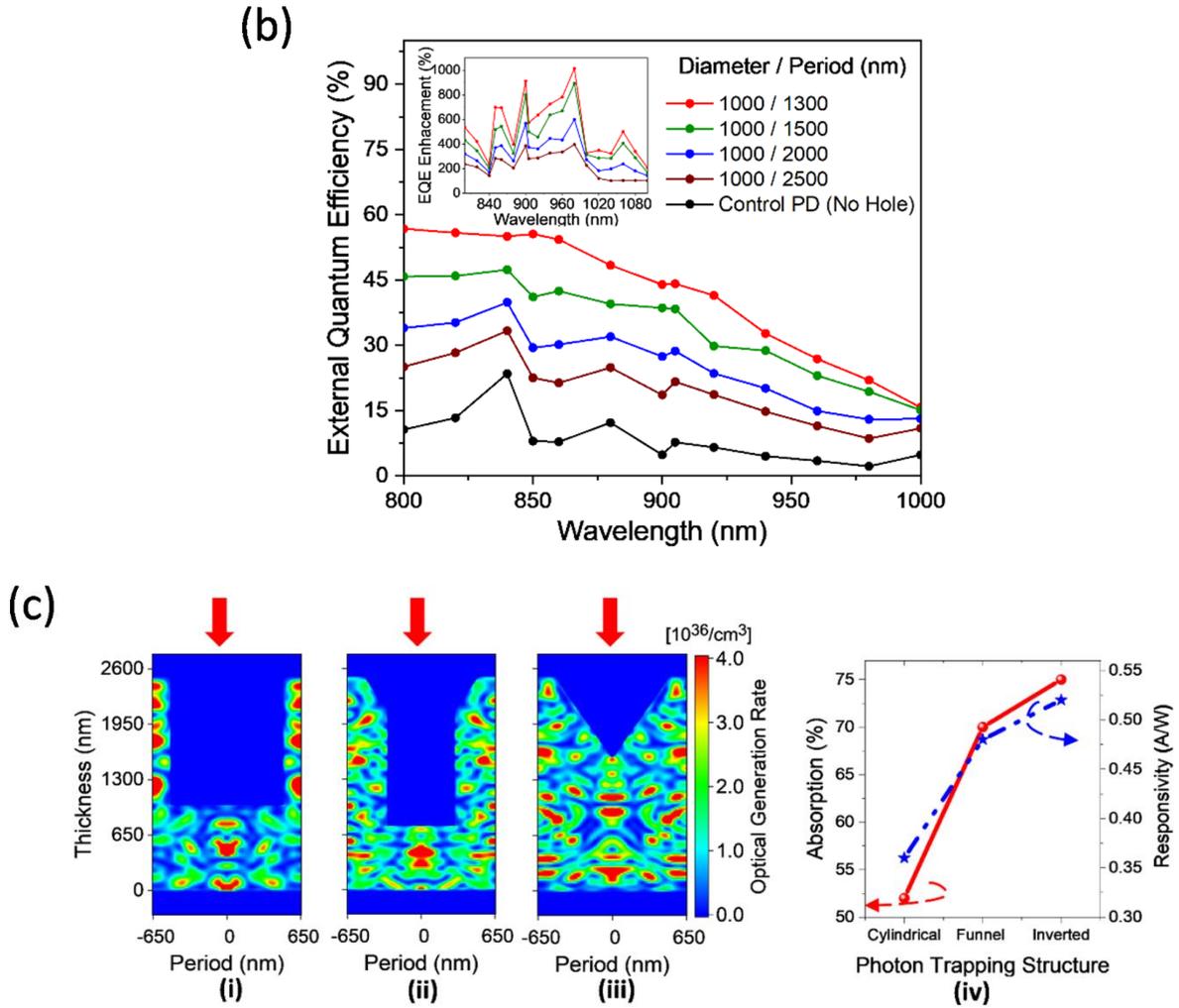

**Figure 2.** a) Experimentally measured EQE of fabricated Si PDs versus nanohole diameter/period for PT structures at 850 nm wavelength, while the PT structures (funnel shape and inverted pyramid) and substrate (bulk Si and SOI) were systematically varied. FDTD simulated EQE of a PD with funnel shape structure is also included (hollow red circle). Devices exhibit EQE from 15% to more than 60%. b) Broadband EQE enhancement in the fabricated PDs for wavelengths ranging from 800 to 1000 nm with a diameter of 1000 nm and decreasing periodicity. Inset: A 10x enhancement in EQE is observed at some wavelengths. c) Optical generation of carriers for different PT profiles: (i) cylindrical, (ii) funnel shape, and (iii) inverted pyramid structures in Si PDs for a monochromatic wavelength of 850 nm. (iv) Calculated maximum photon absorption and responsivity for such PT structures.

Figure 2b illustrates broadband measurements of 500 µm PDs incorporated with inverted pyramids in a hexagonal lattice formation. The wavelengths range from 800 to 1000 nm, while the nanohole p is varied by fixing the d to 1000 nm. Devices with 1000 nm of d and 1300 nm of p pronounce the highest EQE for all the incident wavelengths. This confirms that a high EQE can be attained for a relatively large d/p. Particularly, the maximum efficiencies at wavelengths of 800, 850, and 900 nm are measured as 58%, 56%, and 45%, respectively for the PT devices with $d/p \approx 0.77$. The enhanced absorption coefficients at wavelengths, $\lambda =$ 800, 850, and 900 nm are calculated to be 4335.5, 4104.9, and 2989.2 $cm^{-1}$ by assuming 2 µm of Si i-layer thickness, whereas the absorption coefficients for bulk Si at those λ points are 850, 535, and 306 $cm^{-1}$, respectively. Hence, a maximum of about >10 times higher absorption enhancement is attained at some of the incident wavelengths by integrating inverted pyramids or funnel shape nanoholes in the PDs fabricated on SOI substrates (Figure 2b, inset). Herein, the $SiO_2$ layer of the SOI substrate acts as a mirror due to the high refractive index difference between Si and $SiO_2$, resulting in an enhanced reflection and consequently a higher absorption in the i-layer of the PDs. Furthermore, the optics of the PDs with such PT structures are simulated for the optical generation rate as shown in Figure 2c. PD with cylindrical shape structure is also presented as a comparison. A higher number of modes is seen in inverted pyramid nanohole devices (iii) as compared with cylindrical (i) and funnel shape (ii) nanohole devices due to the reflection at the interface of $SiO_2$ and Si and enhanced lateral propagation of incident light within the devices. Hence, inverted pyramid PDs fabricated on SOI substrates noticeably exhibit higher absorption in comparison with the PDs fabricated on bulk Si and SOI substrates. The simulated inverted pyramid structures exhibit 75% and 0.53 A/W absorption efficiency and responsivity, respectively, while the PDs with etching profiles of the funnel and cylindrical shapes exhibit absorption efficiencies of 70% and 52% and responsivities of 0.49 and 0.36 A/W, respectively (Figure 2c (iv) and S6).

## 2.2. Bandwidth enhancement External

In junction PDs, the 3dB bandwidth is dependent on two parameters: the carrier transit time ($t_r$) and the RC constant-time.[38] A reduction in the junction capacitance due to the presence of the PT nanoholes can be taken into consideration to write the following modified expression of $f_{3dB}$.

$$f_{3dB} = \frac{1}{\sqrt{[2\pi R \times C(1-ff)]^2 + [t_r/0.44]^2}} \quad (3)$$

Where $t_r$ is the transit time required for the carriers to reach the electrode at saturation velocity, $R$ is assumed as 50 Ω, and $ff$ is the filling fraction of the nanohole array (see Figure S8, Supporting Information). By considering the pin PD as a parallel plate, the capacitance can be written as $C=\varepsilon_o\varepsilon_r A/w$, where $\varepsilon_o$ and $\varepsilon_r$ are the permittivity of vacuum and silicon, respectively; $w$ is the depletion layer width, typically the i-layer, and $A$ is the junction area. The integration of an array of PT holes leads to the reduction of effective cross-section area and active materials in the PDs. Consequently, the overall junction capacitance of a PD is reduced proportionally to the ff of the nanohole array. The use of a thin i-layer in conventional PDs reduces the transit time but increases the junction capacitance. Such a trade-off is effectively addressed with integrated PT nanoholes.

**Figures 3**a and b show the results of capacitance-voltage (C-V) measurements performed on the PDs with a diameter of 30 and 80 µm, respectively. The top left insets represent the top view of SEM images of PDs. The experimental C-V measurements between the control and the PT PD show a 15% and 35% of capacitance reduction for PDs with 30 and 80 µm of diameter, respectively. Such reduction is corroborated by applying analytical modeling of the capacitance based on the doping profile and built-in potential as described in Equation S3 of Supporting Information and other references.[39] Furthermore, the same investigation is also conducted for devices with 40 and 50 µm diameter and included in Figure S7. Higher

capacitance reduction is observed as the diameter of the PDs increases since, in our current design, a larger diameter of PD allows to accommodate a higher number of nanoholes. Both devices can reach >50% (see Figure S9, Supporting Information) of capacitance reduction by decreasing the area occupied by the ohmic contacts on the surface of the PDs using CMOS foundries where the width of metal contacts can be less than 150 nm.[40]

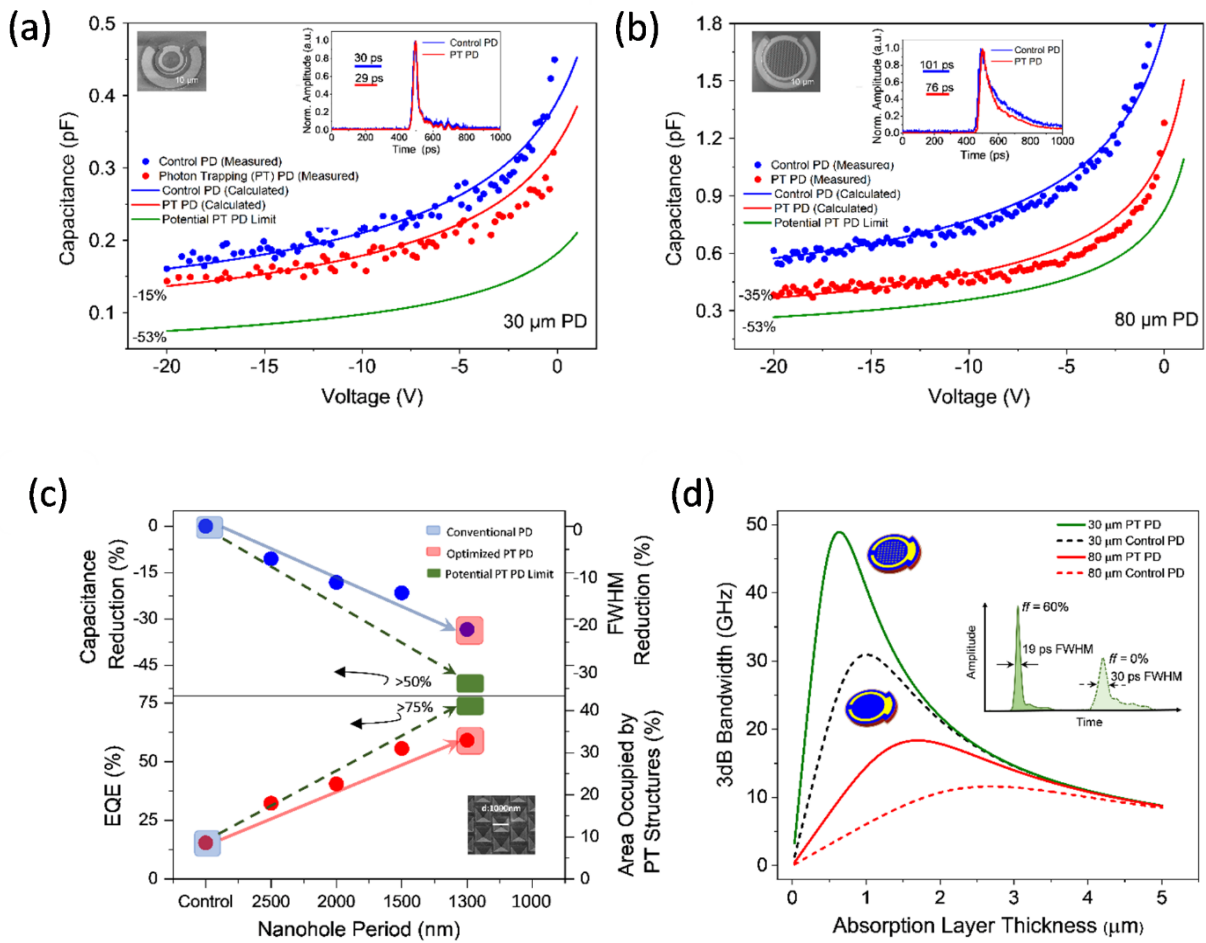

**Figure 3.** Capacitance-voltage characteristics of PDs comparing PT and control device with a) 30 and b) 80 µm diameters, confirming a 15% and 35% capacitance reduction, respectively. This leads to up to 25% narrower FHWM in the pulse time response (inset). Over 50% capacitance reduction can be realized by increasing the number of PT nanoholes. c) A study of >150 different devices are used to optimize PT PDs with simultaneous improvement in EQE, reduction in capacitance, and enhancement in time response. A set of devices with fixed d of 1000 nm and different periods are used to show that >50% of capacitance reduction and >75% of EQE can be achieved at 850 nm. d) Modeling of 3dB bandwidth versus absorption layer thickness considering 60% of capacitance reduction in PT PDs. Illustrations in the inset show a stronger signal amplitude and narrower impulse response of 19 ps is possible in a 30 µm PT PD depicted in Figure 3 a).

The impulse responses of the fabricated PDs are measured under 850 nm illumination and shown in the insets of Figure 3a and 3b. The Full-Width-Half-Maximum (FWHM) of the impulse response has been reduced up to 25% in the PD with 80 µm diameter. This is due to the reduced effective capacitance and consequently reduced RC time in PT PDs compared to the control devices. Higher FWHM and RC time reduction can be achieved in optimally designed PT PDs by fabricating them with closely packed nanoholes and narrower ohmic contacts in advanced semiconductor foundries. The impulse responses of control and PT PDs with larger diameters of 40 and 50 µm are presented in Supporting Information, Figure S7. The collective absorption enhancement of >75 %, capacitance reduction of >50%, and FWHM reduction of > 35% shown in Figure 3c and S9 allow any designer to optimize the PDs with integrated photon-trapping structures. Figure 3d shows how a drastic reduction in the capacitance can dramatically enhance the ultrafast operation of a PD. For example, the impulse response of our control device with 30 µm of diameter is measured to be 30 ps and with optimum PT nanoholes, it can exhibit ~19 ps FWHM (Figure 3d, inset).

## 3. Conclusion

Through extensive simulations and experimental implementations of photon-trapping structures in silicon photodetectors, we helped divulge a direct correlation between the enhancement of absorption and physical parameters of the photon-trapping structures integrated in the photodetectors. We employed cylindrical nanoholes, inverted pyramids, and funnel-shaped surface formations and achieved up to 1000% higher quantum efficiency compared to the control devices. This was made possible by bending the incident beam of light and enabling lateral propagation of modes to prolong the light-matter interactions and suppress back reflection. The enhancement in absorption also comes with a considerable reduction in device capacitance by more than 50% and thereby an improvement in the time response. The combined effect collectively helped to overcome the trade-off between the efficiency and bandwidth of PDs. State-of-the-art CMOS fabrication processes could enable near-perfect EQE and above the 50% of capacitance reduction by increasing the number of photo-trapping nanoholes integrated into the devices (Figure 3c and S8). Analytic equations based on empirical modeling are presented to make it possible to correlate, with high accuracy, the photon-trapping efficiency of the photodetector with physical properties of the photon-trapping structures, material characteristics, and limitations of the fabrication technologies. Such results open opportunities for the development of complete CMOS integrated receivers operating with high sensitivity and high speed and can be expanded to other semiconductors such as germanium (Ge), gallium arsenide (GaAs), and indium phosphide (InP) based ternary and quaternary materials.

## 4. Experimental Section

*Fabrication of photon trapping photodetectors:* Mesa type Si pin PDs are fabricated with a total thickness of 2.5 µm. Heavily doped regions of p and n exhibit reduced electron and hole lifetime of carriers, respectively, facilitating relatively lower diffusion of photogenerated carriers into the high-field i-layer region. Additionally, it reduces the series resistance. The fabrication processes were done in class 100 cleanroom. The PIN wafers on bulk Si or SOI were cleaned in a piranha solution (10:1) to remove organic residues. For inverted pyramids nanoholes, 200 nm of silicon nitride was deposited by PECVD at 250 °C and serves as a masking layer for KOH etching. Next, DUV lithography was used to pattern the nanoholes on the wafer followed by DRIE etching to pattern the silicon nitride. KOH etching was performed with a solution at 33% for 2:30 minutes at 65 °C. Next, DRIE was used to reach the n-mesa and p-mesa of the device, allowing to deposit by evaporation 100 nm of Al and 200nm of Pt that serves as the n-ohmic and p-ohmic contacts. Finally, the wafers were treatment with HF for 10 s to minimize the leakage current. For the fabrication of funnel shape nanoholes [41] and other passivation methods[42] can be found in other references.

*Optical simulation method*: Finite-Difference Time-Domain (FDTD) optical wave propagation simulation method is used to calculate the electromagnetic field distribution within photodetector. In this simulation, a plane wave with a wavelength of 850 nm is a normal incident to the surface of a photodetector. Periodic Boundary Conditions (PBC) are assumed laterally between unit cells and Perfect Matching Layer (PML) boundary conditions are set at the top and bottom of the Si PD. In the first step, the photon absorption is calculated, where electromagnetic field distributions are used as input parameters. The absorption (A) is obtained from the subtraction of the transmission (T) and the reflection (R) as A=1-T-R. In the next, the EQE, which is defined as the ratio of the total incident power on the photodetector to the amount of photon absorbed in the intrinsic layer is calculated, while it is assumed that all the photogenerated carriers are collected by the electrodes. Finally, the current density is calculated from the quantum efficiency.

*Pulse time response measurements:* A mode-locked pulsed laser with a sub-picosecond pulse width is used, where the PDs are illuminated by an 850 nm wavelength with a biasing over 3 V bias. The generated electric pulses are delivered to a 20 GHz sampling oscilloscope via a GSG probe and 25 GHz bias-T

**Supporting Information**

Supporting Information is available from the Wiley Online Library or from the author.

**Acknowledgments**

This work was supported in part by the US Army's Night Vision and Electronic Sensors Directorate under Grant # W909MY-12-D-0008, NSF ECCS grant # 1428392, and by the S. P. Wang and S. Y. Wang Partnership, Los Altos, CA. C.B.P. acknowledges the National Council of Science and Technology (CONACYT) and UC-MEXUS for the Doctoral fellowship provided.

**Competing interests**

The authors declare no competing interests.

**References**

[1]    M.B. Johnston, *Nature Photonics* (**2017**), *11*, 268.

[2]    M.K. Emsley, O. Dosunmu, M.S. Unlu, *IEEE Photonics Technology Letters* (**2002**), *14*, 519.

[3]    S. Riazimehr, S. Kataria, J.M. Gonzalez-Medina, S. Wagner, M. Shaygan, S. Suckow, F.G. Ruiz, O. Engström, A. Godoy, M.C. Lemme, *ACS Photonics* (**2019**), *6*, 107.

[4]    K. Kim, S. Yoon, M. Seo, S. Lee, H. Cho, M. Meyyappan, C.-K. Baek, *Nature Electronics* (**2019**), *2*, 572.

[5]    S. Kim, S. Han, B. Kang, K. Lee, J.D.K. Kim, C. Kim, *IEEE Electron Device Letters* (**2010**), *3*, 1272.

[6]    Y. Kato, T. Sano, Y. Moriyama, S. Maeda, T. Yamazaki, A. Nose, K. Shiina, Y. Yasu, W.v.d. Tempel, A. Ercan, Y. Ebiko, D.V. Nieuwenhove, S. Sukegawa, *IEEE Journal of Solid-State Circuits* (**2018**), *53*, 1071.

[7] D. Stoppa, L. Pancheri, M. Scandiuzzo, L. Gonzo, G.D. Betta, A. Simoni, *IEEE Transactions on Circuits and Systems I: Regular Papers* (**2007**), *54*, 4.

[8] I. Takai, H. Matsubara, M. Soga, M. Ohta, M. Ogawa, T. Yamashita, *Sensors (Basel)* (**2016**), *16*, 459.

[9] L. Zhang, D. Chitnis, H. Chun, S. Rajbhandari, G. Faulkner, D. O'Brien, S. Collins, *Journal of Lightwave Technology* (**2018**), *36*, 2435.

[10] L. Zhang, H. Chun, Z. Ahmed, G. Faulkner, D. O'Brien, S. Collins, *Journal of Lightwave Technology* (**2019**), *37*, 4367.

[11] H. Li, P. Wolf, P. Moser, G. Larisch, J.A. Lott, D. Bimberg, *IEEE Journal of Selected Topics in Quantum Electronics* (**2015**), *21*, 405.

[12] Z. Yang, T. Albrow-Owen, H. Cui, J. Alexander-Webber, F. Gu, X. Wang, T.-C. Wu, M. Zhuge, C. Williams, P. Wang, A.V. Zayats, W. Cai, L. Dai, S. Hofmann, M. Overend, L. Tong, Q. Yang, Z. Sun, T. Hasan, *Science* (**2019**), *365*, 1017.

[13] M.F. Santangelo, E.L. Sciuto, A.C. Busacca, S. Petralia, S. Conoci, S. Libertino, *Sensing and Bio-Sensing Research* (**2015**), *6*, 95.

[14] C. Bruschini, H. Homulle, I.M. Antolovic, S. Burri, E. Charbon, *Light: Science & Applications* (**2019**), *8*, 87.

[15] G. Ariño-Estrada, G.S. Mitchell, S.I. Kwon, J. Du, H. Kim, L.J. Cirignano, K.S. Shah, S.R. Cherry, *Physics in Medicine & Biology* (**2018**), *63*, 04LT01.

[16] H. Savin, P. Repo, G. von Gastrow, P. Ortega, E. Calle, M. Garín, R. Alcubilla, *Nature Nanotechnology* (**2015**), *10*, 624.

[17] E. Garnett, P. Yang, *Nano letters* (**2010**), *10*, 1082.

[18] F. Priolo, T. Gregorkiewicz, M. Galli, T.F. Krauss, *Nat Nanotechnol* (**2014**), *9*, 19.

[19] S.F. Leung, Q. Zhang, F. Xiu, D. Yu, J.C. Ho, D. Li, Z. Fan, *J Phys Chem Lett* (**2014**), *5*, 1479.

[20] M.L. Brongersma, Y. Cui, S. Fan, *Nat Mater* (**2014**), *13*, 451.

[21]     C. Lin, M.L. Povinelli, *Opt. Express* **(2009)**, *17*, 19371.

[22]     L.-B. Luo, L.-H. Zeng, C. Xie, Y.-Q. Yu, F.-X. Liang, C.-Y. Wu, L. Wang, J.-G. Hu, *Scientific Reports* **(2014)**, *4*, 3914.

[23]     V.E. Ferry, L.A. Sweatlock, D. Pacifici, H.A. Atwater, *Nano Letters* **(2008)**, *8*, 4391.

[24]     Y. Gao, H. Cansizoglu, K.G. Polat, S. Ghandiparsi, A. Kaya, H.H. Mamtaz, A.S. Mayet, Y. Wang, X. Zhang, T. Yamada, E.P. Devine, A.F. Elrefaie, S.-Y. Wang, M.S. Islam, *Nat Photon* **(2017)**, *11*, 301.

[25]     K. Zang, X. Jiang, Y. Huo, X. Ding, M. Morea, X. Chen, C.Y. Lu, J. Ma, M. Zhou, Z. Xia, Z. Yu, T.I. Kamins, Q. Zhang, J.S. Harris, *Nat Commun* **(2017)**, *8*, 628.

[26]     H. Cansizoglu, C. Bartolo-Perez, Y. Gao, E.P. Devine, S. Ghandiparsi, K.G. Polat, H.H. Mamtaz, T. Yamada, A.F. Elrefaie, S.-Y. Wang, *Photonics Research* **(2018)**, *6*, 734.

[27]     M.-L. Hsieh, A. Kaiser, S. Bhattacharya, S. John, S.-Y. Lin, *Scientific Reports* **(2020)**, *10*, 11857.

[28]     Z. Yu, A. Raman, S. Fan, *Proc Natl Acad Sci U S A* **(2010)**, *107*, 17491.

[29]     H. Cansizoglu, A.S. Mayet, S. Ghandiparsi, Y. Gao, C. Bartolo-Perez, H.H. Mamtaz, E.P. Devine, T. Yamada, A.F. Elrefaie, S. Wang, M.S. Islam, *IEEE Photonics Technology Letters* **(2019)**, *31*, 1619.

[30]     L. Frey, M. Marty, S. Andre, N. Moussy, *IEEE Journal of the Electron Devices Society* **(2018)**, *6*, 392.

[31]     S. Ghandiparsi, A.F. Elrefaie, A.S. Mayet, T. Landolsi, C.Bartolo-Perez, H. Cansizoglu, Y. Gao, H. Mamtaz, H.R. Golgir, E.P. Devine, T. Yamada, S. Wang, M.S. Islam, *Journal of Lightwave Technology* **(2019)**, 37, 5748.

[32]     X. Cao, Y. Zhang, Z. Han, W. Li, Y. Jin, A. Wu, *Journal of Materials Science: Materials in Electronics* **(2020)**, *31*, 5872.

[33]     H. Zhou, S. Xu, Y. Lin, Y.-C. Huang, B. Son, Q. Chen, X. Guo, K.H. Lee, S.C.-K. Goh, X. Gong, C.S. Tan, *Opt. Express* **(2020)**, *28*, 10280.

[34]	S. Baruah, J. Bora, S. Maity, *Microsystem Technologies*, **(2020)**.

[35]	 D. Chen, K. Sun, A.H. Jones, J.C. Campbell, *Opt. Express* **(2020)**, *28*, 24379.

[36]	P. Kuang, S. Eyderman, M.L. Hsieh, A. Post, S. John, S.Y. Lin, *ACS Nano* **(2016)**, *10*, 6116.

[37]	J. Zhao, M.A. Green, *IEEE Transactions on Electron Devices* **(1991)**, *38*, 1925.

[38]	S.B. Alexander, *Optical communication receiver design*, SPIE Press, **1997**.

[39]	 K.K. Ng, Sze, S. M., Physics of semiconductor devices, in: *Physics of Semiconductor Devices, John Wiley & Sons, Inc*, **2006**.

[40]	W. Zhi, Q. Quan, P. Yu, Y. Jiang, *Micromachines (Basel)* **(2020)**, *11*, 65.

[41]	Y. Gao, H. Cansizoglu, S. Ghandiparsi, C. Bartolo-Perez, E.P. Devine, T. Yamada, A.F. Elrefaie, S.-y. Wang, M.S. Islam, *ACS Photonics* **(2017)**.

[42]	A.S. Mayet, H. Cansizoglu, Y. Gao, S. Ghandiparsi, A. Kaya, C. Bartolo-Perez, B. AlHalaili, T. Yamada, E.P. Devine, A.F. Elrefaie, *JOSA B* **(2018)**, *35*, 1059.